\documentclass{PoS}

\usepackage[authoryear,square]{natbib}
\bibpunct{(}{)}{;}{a}{}{,}

\usepackage{textcmds}

\usepackage{aas_macros}

\newcommand{\url}[1]{\href{#1}{\tt #1}}

\title{Magnetic fields on a wide range of scales in star-forming galaxies}

\ShortTitle{Magnetic fields in galaxies}

\author{\speaker{George Heald},$^{abc}$ Anna Williams$^{d}$ and Sarrvesh S. Sridhar$^{cb}$\\
        \llap{$^a$}CSIRO Astronomy and Space Science, 26 Dick Perry Avenue, Kensington WA 6151, Australia\\
        \llap{$^b$}ASTRON, the Netherlands Institute for Radio Astronomy, Postbus 2, 7990 AA, Dwingeloo, The Netherlands\\
        \llap{$^c$}Kapteyn Astronomical Institute, University of Groningen, PO Box 800, 9700 AV, Groningen, The Netherlands\\
        \llap{$^d$}Department of Astronomy, University of Wisconsin-Madison, 475 N. Charter Street, Madison, WI, USA\\
        E-mail: \email{George.Heald@csiro.au}}

\abstract{
A key ingredient in the evolution of galaxies is the star formation cycle. Recent progress in the study of magnetic fields is revealing the close connection between star formation and its effect on the small-scale structure in the magnetized interstellar medium (ISM).
In this contribution
we describe how the modern generation of radio telescopes is being used to probe the physics of the ISM through sensitive multiwavelength surveys of gas and magnetic fields, from the inner star forming disk and outward into the galaxy outskirts where large-scale magnetic fields may also play a key role. We highlight unique pioneering efforts towards performing and scientifically exploiting large-scale surveys of the type that the SKA will undertake routinely. Looking to the future, we describe plans for using the Square Kilometre Array (SKA) and its pathfinders to gain important new insights into the cosmic history of galaxy evolution.
}

\FullConference{EXTRA-RADSUR2015 (*)\\
		20--23 October 2015\\
		Bologna, Italy

                \bigskip
                \hrule
                \bigskip

                \textnormal{(*) This conference has been organized
                  with the support of the Ministry of Foreign Affairs
                  and International Cooperation, Directorate General
                  for the Country Promotion (Bilateral Grant Agreement
                  ZA14GR02 - Mapping the Universe on the Pathway to
                  SKA)}
}

\newcommand{\HI}{{\sc H\,i}}
\newcommand{\HII}{{\sc H\,ii}}

\begin{document}

\section{Introduction}\label{section:intro}

Magnetic fields are a crucial element of galaxies, and play an important role in setting the stage for their structure and evolution. At the scale of active star formation, the role of magnetism is becoming clear through simulations \citep[e.g.,][]{birnboim_etal_2015,kim_ostriker_2015} and novel observations \citep{planck_2016}. Simulations are also key to understanding the importance of magnetism in defining the multiphase structure of the interstellar medium \citep[ISM; e.g.,][]{hill_etal_2012}. On larger scales, it is still unclear to what degree magnetic forces play an important dynamical role \citep[see, e.g.,][and references therein]{elstner_etal_2014}, but magnetic field lines are certainly essential for guiding the propagation of cosmic rays \citep[see][]{kotera_olinto_2011} and thus in understanding the isotropy of ultra-high energy cosmic ray flux as seen from the Earth \citep[e.g.,][]{alvarez_etal_2002}. Magnetic fields may contribute to setting the properties of galactic outflows \citep[see][]{heesen_etal_2011} and the disk-halo interface \citep[e.g.,][]{benjamin_2002,henriksen_irwin_2016}, and on the largest scales to connecting galaxies to the intergalactic medium \citep[IGM; e.g.][]{kronberg_etal_1999,bernet_etal_2013}. A clear understanding of the properties (shape, strength, and energetic content) of galactic magnetic fields on all scales is plainly of great astrophysical interest.

Much has been written about the ``classical'' techniques for detecting and characterising magnetic fields in galaxies. See for example \citet{condon_1992,heiles_robishaw_2009,andersson_2015,heald_2015,beck_2016} and references therein. In this contribution, we focus on the detailed study of synchrotron radiation and its polarization, although optical/infrared polarization and Zeeman splitting also provide extremely useful and unique information about galactic magnetic fields. Radio synchrotron measurements can be used as powerful and highly flexible probes of various aspects of the ordered and turbulent components of the magnetic field. It is important to distinguish the ordered/regular components of the field from the turbulent/tangled components. These have different observational signatures. For an excellent summary and illustration, see \citet{jaffe_etal_2010}. Crucially, total synchrotron intensity measurements probe the total magnetic field (ordered plus disordered components), while the corresponding polarized emission probes only the ordered (but not necessarily uniformly directed) fields in the plane of the sky; the Faraday rotation experienced by that polarized emission yields estimates of the uniformly directed (regular) field along the line of sight. Polarization is particularly powerful in distinguishing various subtly different configurations of gas, magnetic fields, and cosmic rays within the telescope beam as well as across extended sources. It is in fully understanding and exploiting these detailed measurements that defines the modern era of radio polarimetry. Depolarization, often viewed as a hinderance, is uniquely sensitive to the magnetic field structure through its strong $\lambda^2$ dependence, and provides a detailed probe of magnetism \citep[see, e.g.,][]{horellou_fletcher_2014}.

A basic but powerful tool that is often used to try to make sense of broadband polarization measurements is the Rotation Measure (RM) Synthesis technique \citep{burn_1966,brentjens_debruyn_2005,heald_2009}. The concept has this year reached its $50^\mathrm{th}$ anniversary, but it has only been widely used in the last decade. The procedure relies on a Fourier-like relationship between the observed linear polarization quantities (Stokes {\it Q} and {\it U}), and the amount of polarized emission at each Faraday depth ($\phi$) --- a quantity that describes how much Faraday rotation has occurred along the line of sight $l$ (specifically, $\phi\propto\int\,n_e\vec{B}\cdot d\vec{l}$). A deconvolution approach can be used to attempt to isolate the true signal from the instrumental sampling function \citep[{\tt RMCLEAN};][]{heald_etal_2009}. However, it has been noted that systematic issues can prevent a reliable reconstruction of the input signal when using these techniques \citep[e.g.,][]{farnsworth_etal_2011}. As a consequence, a complementary technique is generally needed: directly fitting the primary broadband {\it QU} data to theoretical models \citep{farnsworth_etal_2011,osullivan_etal_2012,sun_etal_2015}.
In practice, a combination of these techniques should be used because while RM Synthesis can be used arbitrarily with some important limitations, the appropriate physical model must first be correctly identified for {\it QU} fitting to be successful. Indeed, the output of RM Synthesis and {\tt RMCLEAN} can be used as a prior for running {\it QU} fitting \citep[e.g.,][]{mao_etal_2015}, highlighting the complementary nature of the two analytical approaches.

The key distinguishing features that define various physical models are encoded in the detailed properties of the resulting broadband depolarization.
\begin{itemize}
\item \underline{Beam depolarization}: This is a largely $\lambda$-independent effect. It can of course be addressed by observing with improved angular resolution.
\item \underline{Faraday depolarization}: These effects are all dependent on $\lambda^2$, but each has distinct features provided sufficient frequency coverage is utilized. For example, here is a partial list of simple geometries of relevance for the discussion in \S\,\ref{section:smallscale}:
\begin{itemize}
\item {\it Differential Faraday rotation (DFR)}: Caused by a regular field in a region that is both emitting and rotating. Emission originating from locations closer to the observer experience less Faraday rotation; thus polarized signal is present over a range of Faraday depth.
\item {\it Internal Faraday dispersion (IFD)}: Turbulent and regular field in a region that is both emitting and rotating. The turbulence also generates depolarization in addition to the DFR.
\item {\it External Faraday dispersion or inhomogeneous Faraday screen (IFS)}: The emitting and turbulent rotating regions are separated, and the rotating screen is turbulent. It may only cover part of the emitting source (partial IFS; PIFS).
\item {\it Extended partial coverage model}: Partial coverage of a depolarizing screen as above (i.e., partial IFS or PIFS), as well as a depolarizing screen associated with the emitting source.
\end{itemize} 
\end{itemize}
Modern radio telescopes are very well suited to distinguish and characterise the corresponding astrophysical situations \citep{heald_etal_2015}. This contribution will illustrate that depolarization provides a highly useful tool both to learn about detailed small-scale ISM physics (\S\,\ref{section:smallscale}), and to `dissect' large-scale magnetic structure (\S\,\ref{section:largescale}).

Based on the use of standard techniques, as well as the careful interpretation of earlier narrowband data, a consistent and detailed picture of the broad properties of galactic magnetic fields has developed over the course of the past several decades. The general structure is reviewed by \citet{beck_2016}. Typically, galaxies demonstrate axisymmetric spiral fields in the disk, and X-shaped vertical extensions when viewed from an edge-on perspective. Field strengths vary but are typically in the range of several to tens of $\mu\mathrm{G}$, with generally similar contributions from ordered and random field components. The magnetic energy density in galaxy ISMs can be comparable to, or even greater than, that of other contributors to the energy budget. One of the areas where new progress is now being made is in developing a better picture of the small scales, understanding the detailed internal structure of the magnetized ISM and the connection between magnetic fields, multiphase gas, and star formation. Meanwhile, various observational techniques are beginning to unveil the properties of magnetic fields and gas at the outskirts of galaxies. Still, many of the observations that have been employed so far probe two-dimensional (2D) projections of the magnetic field structure. By embracing the effects of depolarization, multifrequency observations now allow us to peel the onion of galactic magnetic fields and unveil their 3D structure.

\section{Small scale magnetic fields}\label{section:smallscale}

\begin{figure}[b]
\centering
\includegraphics[height=0.5\hsize]{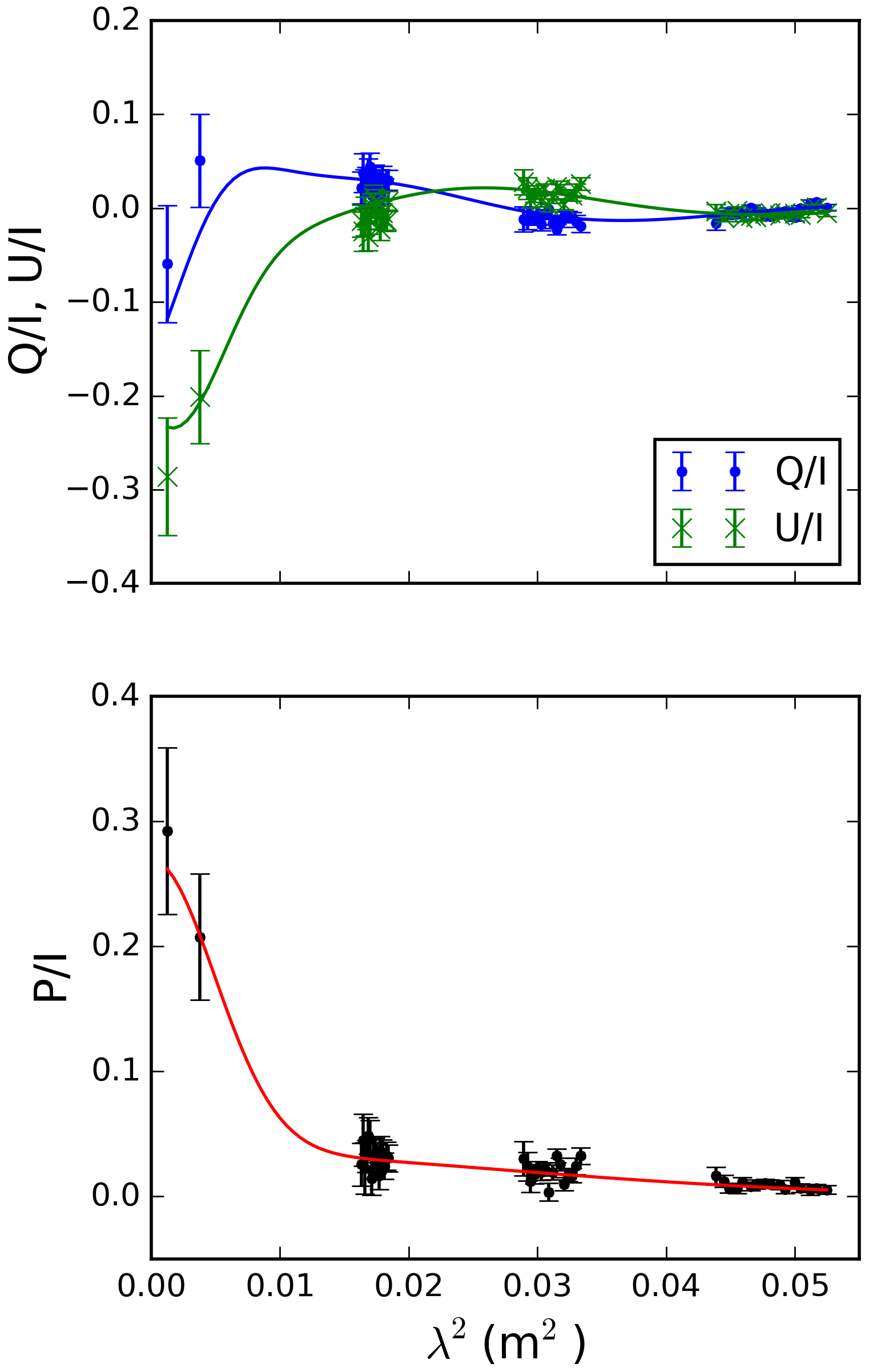}\hfill\includegraphics[height=0.5\hsize]{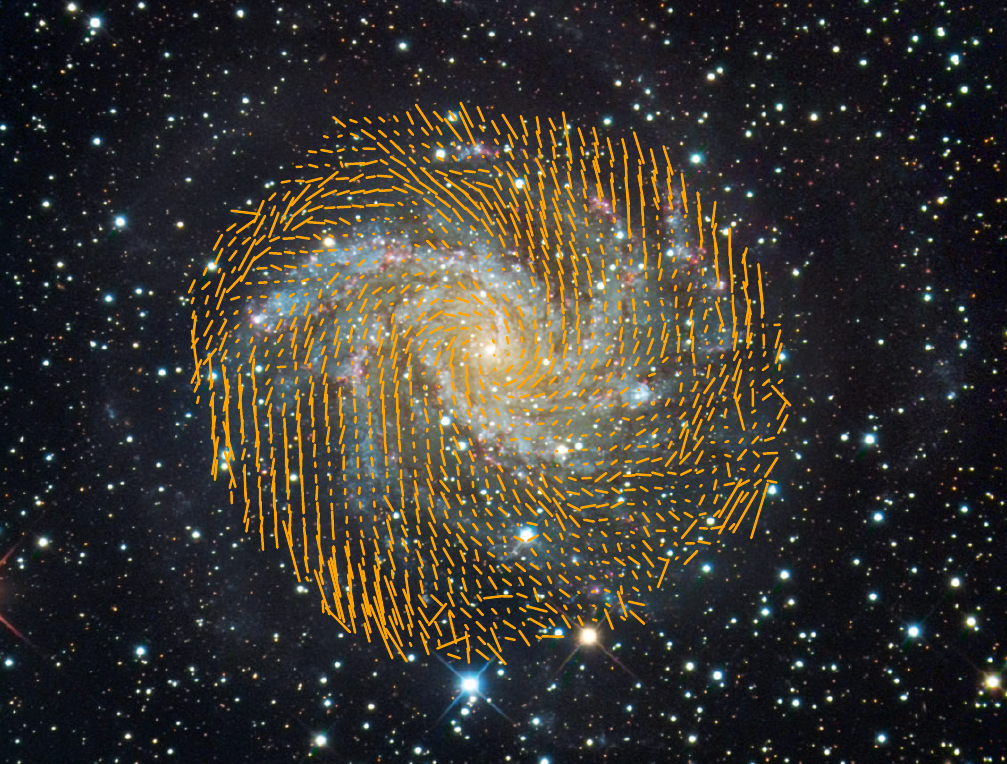}
\caption{Illustration of the depolarization modeling that is revealing the detailed magnetic field structure in NGC~6946 (Williams et al., in prep). Left: Observed fractional polarization values as a function of $\lambda^2$ in a particular region of the galaxy, along with a model fit to the data (in this case, a PIFS model; see \S\,\protect\ref{section:intro}). The top-left panel shows the two linear Stokes parameters {\it Q} and {\it U}, while the bottom-left panel shows $P=\sqrt{Q^2+U^2}$. Right: Reconstruction of the underlying regular field in NGC~6946, recovered after modeling out the depolarization structure in the ISM.}
\label{figure:smallscales}
\end{figure}

It remains unclear how magnetic fields are organized on small scales ($\lesssim1\,\mathrm{kpc}$) within galaxies. The smallest scales can most readily be observed in the Milky Way, where a turbulent spectrum is observed with characteristic scales reminiscent of the size of \HII\ regions and supernova remnants, depending on location in the disk \citep[e.g.,][and references therein]{haverkorn_etal_2008}. A number of observational techniques have been brought to bear in external galaxies, including analysis of RMs from background sources \citep{gaensler_etal_2005} and from the diffuse magnetized ISM itself \citep{mao_etal_2015}, from dispersion of RMs \citep{fletcher_etal_2011} or polarization angles \citep{houde_etal_2013}, and potentially in the future from fluctuation statistics of the synchrotron intensity \citep[total intensity together with polarized intensity;][]{herron_etal_2016}. Still, a comprehensive observational picture of the small-scale structure and relevant spatial scales of interstellar magnetic fields has not yet been developed. What are the typical properties of magnetic field fluctuations across the galaxy population, and what are the relative contributions of ordered and random magnetic field components? To what degree are these related to the structure of the gaseous ISM and the star formation activity taking place in the disk?

Modern broadband polarimetric observations are now allowing a more detailed study of the magnetized ISM in galaxies. An example is illustrated in Figure~\ref{figure:smallscales}. By drawing together polarization maps of NGC~6946 at 3, 6, 13, 18, and 22~cm, Williams et al. (in prep) are able to model the structure of the magnetized ISM using simple configurations of the type summarized in \S\,\ref{section:intro}.
At lower frequencies (i.e., larger values of $\lambda^2$) the data are of sufficiently high spectral resolution and sensitivity that a combination of RM Synthesis and {\it QU} fitting is beginning to distinguish depolarization caused within the emitting medium from the effect of a foreground screen.  Meanwhile, the spatial resolution ($\sim400$~pc) is approaching the scale of giant molecular clouds (GMCs).  New 3D descriptions of the small-scale magnetic fields in external galaxies can now be made.
These modeling efforts not only reveal how the magnetoionized ISM varies across the galaxy, but also allows a reconstruction of the underlying large-scale magnetic field shape, as seen in Figure~\ref{figure:smallscales}.

\section{Large scale magnetic fields}\label{section:largescale}

\begin{figure}
\centering
\includegraphics[width=0.5\hsize]{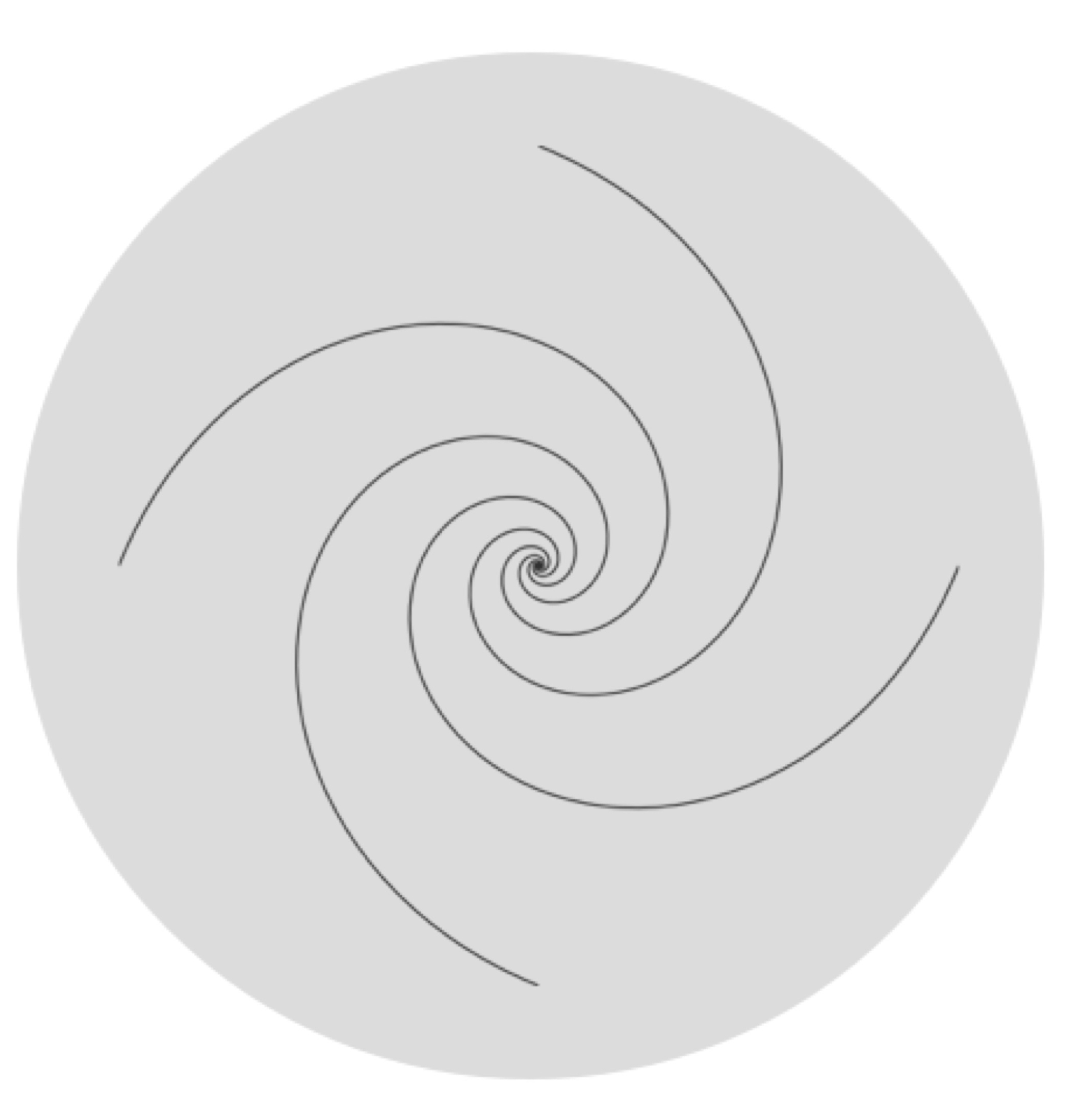}\includegraphics[width=0.5\hsize]{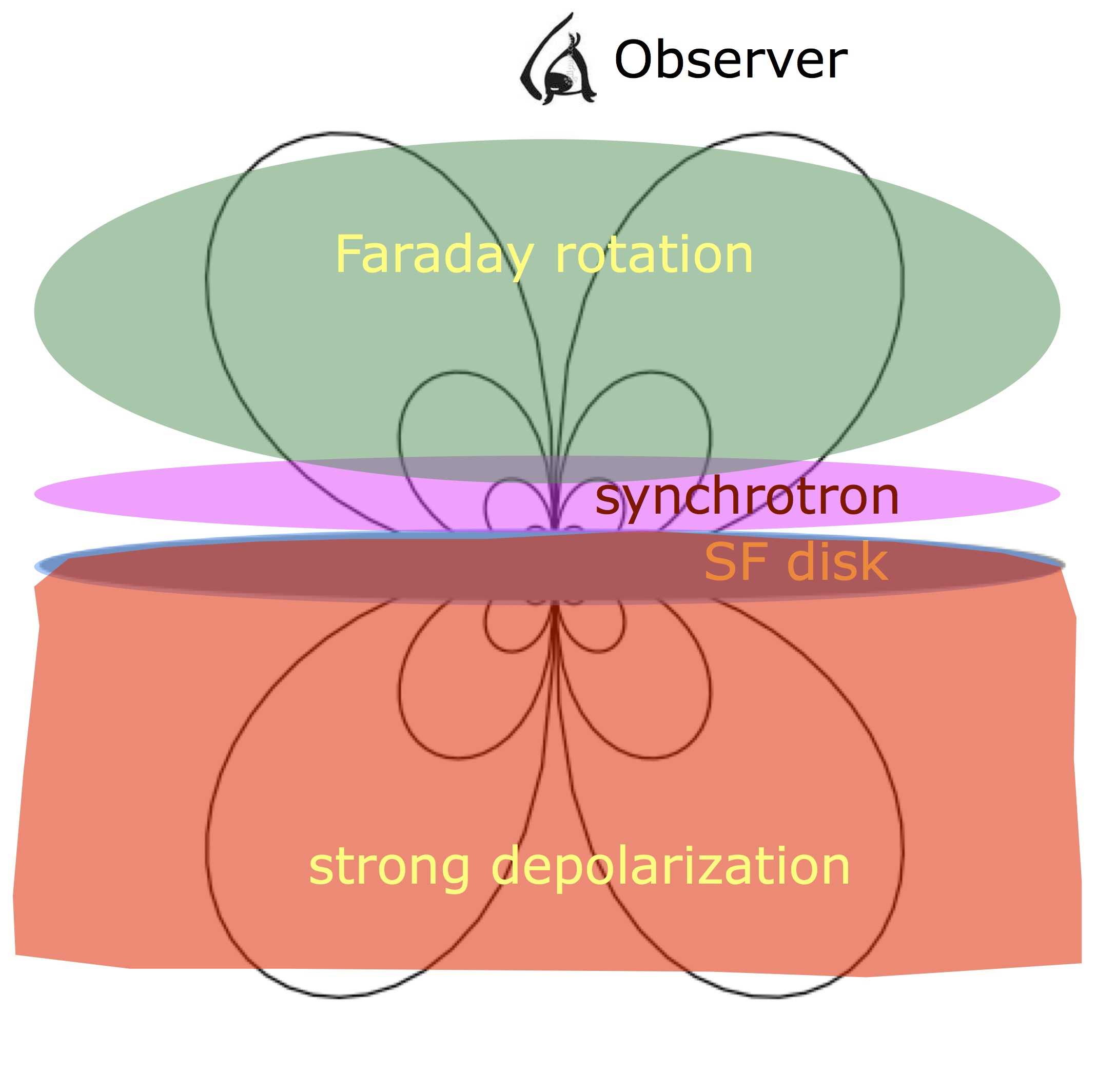}
\caption{Illustration of the magnetic field and depolarization model proposed by \citet{braun_etal_2010}. Left: Axisymmetric spiral field in the disk. Right: Quadrupolar poloidal field as viewed from an edge-on perspective. Due to the turbulent depolarization in the midplane (induced by active star formation injecting energy into the ISM; {\it blue region}), strong depolarization ({\it red region}) prevents the observer from detecting polarization at $\sim1\,\mathrm{GHz}$ within and behind the thin disk. Polarized synchrotron radiation ({\it purple region}) is visible to such an observer from the front surface of the disk only. That polarized signal may be Faraday rotated by thermal electrons and magnetic fields in the halo ({\it green region}).}
\label{figure:asymmetry}
\end{figure}

An interesting development in the past few years has been the use of depolarization as a tool to permit `slicing' galaxies into different depths to investigate the structure of the 3D field in a tomographic manner. For example, \citet{heald_etal_2009} and \citet{braun_etal_2010} have used this approach to first recognize and then interpret generally applicable, large-scale symmetric patterns in the distribution of polarized synchrotron emission from a large number of galaxy disks. Specifically, they noted that for moderately inclined galaxies, a minimum in polarization consistently occurs along the kinematically receding major axis. This feature has been subsequently recognized in other galaxy samples \citep[e.g.,][]{vollmer_etal_2013}. It is illustrated in Figure~\ref{figure:asymmetry}. Along with an azimuthal variation in polarized emissivity, the galaxies also demonstrate a corresponding azimuthal variation in the Faraday rotation measure. Finally, \citet{braun_etal_2010} report `ghost' polarized emission in some galaxies, at extreme values of RM \citep[but cf.][]{mao_etal_2015}.

The cause of these polarized features is interpreted in the following way. Turbulent depolarization in the midplane, induced by active star formation, serves to depolarize synchrotron radiation from within the disk and to a large degree from the back-side thick disk region as well. (Ghost emission at extreme RM, if present, is interpreted to originate from the back-side of the disk.) In this picture, at $\sim1~\mathrm{GHz}$ frequency, the bulk of the polarized emission originates from a layer just above the star forming thin disk \citep[see also][]{horellou_etal_1992}. Due to a combination of azimuthal and vertical magnetic field components, the observed polarized asymmetry is expected to be missing from the region where the ordered magnetic field points predominantly toward the observer. \citet{braun_etal_2010} model simple magnetic field configurations and conclude that the observations are consistent with a mixture of an axisymmetric spiral field and a quadrupolar poloidal field. This picture works if galaxies have trailing spiral arms, as indeed seems to be the case \citep{devaucouleurs_1958}. Galaxies with leading arms would be expected to have a minimum in polarization along the approaching major axis. A further prediction of the model is that the asymmetry should vanish at higher frequency due to the lower impact of turbulent depolarization, as is indeed seen in the case of NGC~6946 \citep{beck_2007} at wavelengths of 3 and 6~cm.

A rather more detailed application of the same basic analysis has been performed for the individual case of M~51 \citep{fletcher_etal_2011}. By drawing together a large multifrequency data set spanning wavelengths from 3 to 20~cm, the authors were able to consider separately the thermal electron population and magnetic field properties in the disk and halo regions. Ultimately they found that different spiral modes were present in the magnetic fields of the disk and halo ($m=0+2$ in the disk, and $m=1$ in the halo; note that the qualitative appearance of the RM maps presented by \citet{heald_etal_2009} are consistent both with this model and with the interpretation of the $\sim1\,\mathrm{GHz}$ emission originating from above the disk).

\begin{figure}
\centering
\includegraphics[height=0.425\hsize]{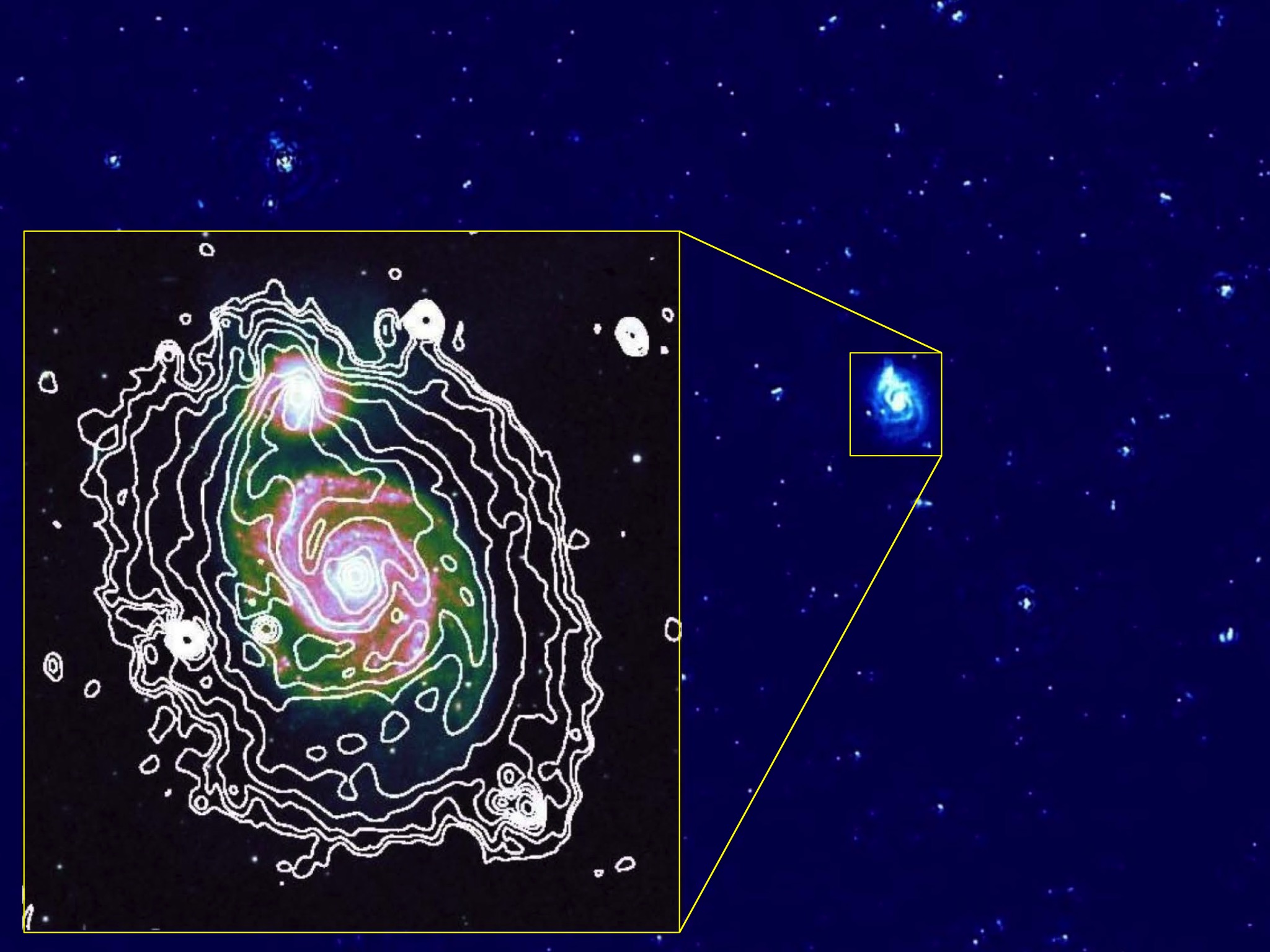}\hfill\includegraphics[height=0.425\hsize]{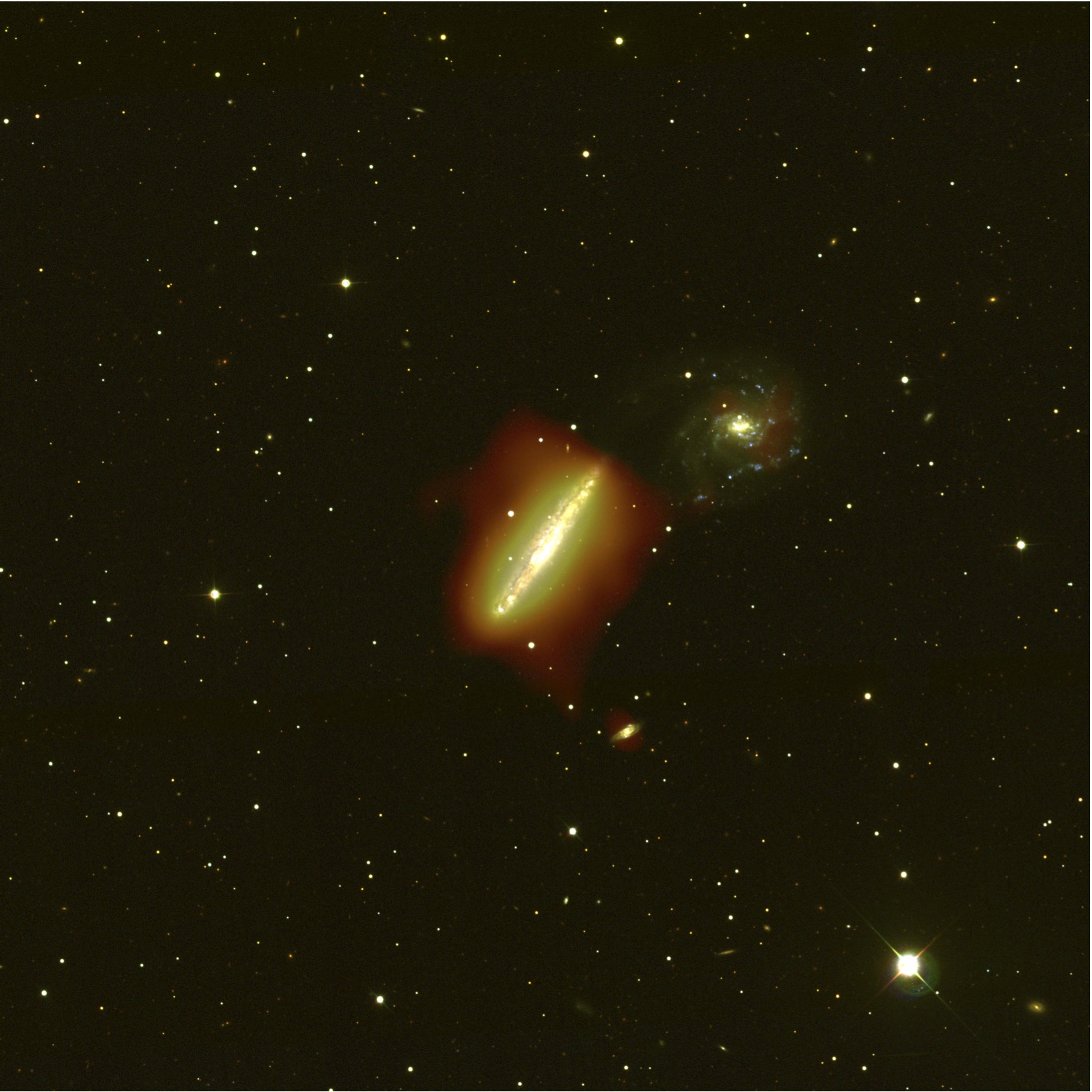}
\caption{Examples of nearby galaxies viewed at low radio frequency (here, $150\,\mathrm{MHz}$) with high angular resolution ($\sim20^{\prime\prime}$) and excellent sensitivity ($<1\,\mathrm{mJy\,beam}^{-1}$) using LOFAR. Left: Wide field of view image surrounding the nearby galaxy M51 \citep{mulcahy_etal_2014}. The inset shows a DSS image of the galaxy overlaid with LOFAR contours, starting at $1\,\mathrm{mJy\,beam}^{-1}$ and increasing by factors of 1.5. Right: False-color SDSS composite of the nearby edge-on galaxy NGC~5775 (Heald et al., in prep) and its vertically extended radio halo (orange colors), visible with LOFAR extending up to 15~kpc from the star forming disk. The companion galaxy NGC~5774 is also detected, as is a bridge of radio emission corresponding to an \HI\ bridge \citep[see][]{irwin_1994}.}
\label{figure:lofar}
\end{figure}

Low-frequency ($\nu\lesssim300\,\mathrm{MHz}$) synchrotron radiation probes even farther into the outskirts of galaxies. While synchrotron and inverse Compton energy losses mean that cosmic rays no longer emit radiation at high frequencies once they have propagated far from regions of star formation, the low-energy cosmic ray population loses energy more slowly \citep[$\dot{E}\propto\,E^2$;][]{condon_1992} and thus emits at low radio frequency for a longer span of time, i.e. $\tau\propto\,E^{-1}$. The new generation of low frequency radio telescopes is in this way opening the window to the outer parts of galactic magnetic fields. For example, total intensity LOFAR observations have been used to probe into the far outer disk \citep[e.g., in M51;][]{mulcahy_etal_2014} and into the upper disk-halo interface region (e.g., NGC 5775; Heald et al. in prep, see Figure \ref{figure:lofar}). One ultimate aim for this kind of research is to start to recover polarization at such large distances from the star forming disk, in order to further constrain the {\it ordered} component of the large-scale field (and possibly also the regular component from associated RMs).

At the very largest distance from the star forming regions of galaxies, where cosmic rays have lost too much energy to be detectable in synchrotron radiation with current technology, different methods must be brought to bear to identify and study the extended magnetic field structure. The Universe provides excellent probes of galactic magnetic fields on the largest scales: background polarized radio galaxies \citep[as have been used to study the magnetic extent of galaxy clusters;][]{clarke_etal_2001}. For galaxies, a similar approach has led \citet{bernet_etal_2013} to find evidence for strong magnetic fields (tens of $\mu$G) extending up to 50~kpc from galaxies, and with properties suggestive of galactic outflows rather than large coherent magnetic structures. Follow-up studies with forthcoming radio surveys will be of particular interest in the next several years.

%

\section{Future prospects}

While recent progress in our understanding of magnetism has been rapid and very encouraging, there is also still clearly much to learn using future radio telescopes and accompanying large-scale survey efforts. For example, key questions remain surrounding the buildup and evolution of magnetism in the Universe and in bound objects like galaxies; the detailed internal structure of magnetic fields on both small and large scales in galaxies; and the interplay between magnetic fields and the gaseous content and kinematics in galaxies. In this section, we provide a brief overview of some of the radio surveys that are currently underway and will be pursued with the next generation of observatories, leading to the SKA.

One interesting survey that is currently enhancing our view of the magnetic content of nearby galaxies is called ``Continuum HAlos in Nearby Galaxies -- an EVLA Survey'' \citep[CHANG-ES;][]{irwin_etal_2012a}. This project makes use of sensitive broadband continuum observations of a large number (35) of edge-on nearby star-forming galaxies, in full polarization, at L- and C-bands ($1.2-1.9$ and $5-7\,\mathrm{GHz}$) using the Jansky Very Large Array (VLA), or EVLA as it was known at the time. From these observations a clear picture of the vertical extent of the magnetic component of nearby galaxies is starting to be revealed \citep[e.g.,][]{irwin_etal_2012b,wiegert_etal_2015}. Surveys of this type are set to become commonplace as improvements in the sensitivity, field of view, and instantaneous bandwidth of radio telescopes are ubiquitous. Two aspects that will need continued care for future efforts are ({\it i}) instrumental polarization, not just at the pointing center but across the whole field of view (requiring control of, and excellent understanding of, the beamshape), and ({\it ii}) characterization of the ionospheric contribution to Faraday rotation measure.

As we look toward the horizon, the community is greatly anticipating the next generation of radio telescope facilities, amongst these the Australian Square Kilometre Array Pathfinder \citep[ASKAP;][]{johnston_etal_2008} and the South African MeerKAT \citep[formerly the Karoo Array Telescope;][]{jonas_2009}. Both of these will incorporate polarimetric observations either as dedicated surveys or as fundamental aspects of larger survey activities. These new telescopes promise to deliver both excellent polarization performance as well as the broad bandwidth that is needed to characterize the internal structure of the astrophysical sources of interest. Particular surveys of note include the Polarization Sky Survey of the Universe's Magnetism \citep[POSSUM;][]{gaensler_2009} with the ASKAP telescope, which will develop a catalog containing an estimated $\sim3$ million polarized sources over a surveyed sky area of $30,000$ square degrees. By covering the frequency range $1150-1450\,\mathrm{MHz}$, accurate RMs will be available for a large number of those sources, leading to an ``RM grid'' of POSSUM sources with a density as high as 100~RMs~sq~deg$^{-1}$. The catalog will allow large statistical studies of polarized source populations, as well as enabling investigations of the extended magnetic structure of intervening objects along the line of sight (e.g. the Milky Way, nearby galaxies and clusters). In South Africa, the MeerKAT \HI\ Observations of Nearby Galactic Objects: Observing Southern Emitters (MHONGOOSE)\footnote{See \url{http://mhongoose.astron.nl/}} survey will probe the magnetic fields in 30 nearby galaxies to exquisite depth (200 hr per target) and with broadband frequency coverage. Meanwhile the MeerKAT International GigaHertz Tiered Extragalactic Exploration (MIGHTEE)\footnote{See \url{http://public.ska.ac.za/meerkat/meerkat-large-survey-projects}} survey will observe broadband continuum emission in a tiered approach, covering 1000 square degrees to a depth of $\sim5\,\mu\mathrm{Jy\,beam}^{-1}$ along with two smaller and deeper sky areas, enabling great strides in understanding the cosmic evolution of magnetism in distant AGN and star-forming galaxies.

On somewhat longer timescales, the community is working toward transformational breakthroughs in several facets of cosmic magnetism science with the SKA. Several headline science projects are being developed, with the highest priority embodied in a large-area, deep, broadband survey for Faraday rotation measures \citep{johnston-hollitt_etal_2015}. Additional science cases include those focused on nearby galaxies \citep{heald_etal_2015,beck_etal_2015}, the Milky Way galaxy \citep{haverkorn_etal_2015}, detailed broadband studies of radio galaxies \citep{gaensler_etal_2015}, and intermediate-redshift normal and starforming galaxies \citep{taylor_etal_2015}. There clearly remain a great number of outstanding questions of fundamental importance that will be pursued in the coming decades of radio surveys.

\acknowledgments{LOFAR, the Low Frequency Array designed and constructed by ASTRON, has facilities in several countries, that are owned by various parties (each with their own funding sources), and that are collectively operated by the International LOFAR Telescope (ILT) foundation under a joint scientific policy.}

\end{document}